\documentclass{article}

\usepackage{PRIMEarxiv}

\usepackage[utf8]{inputenc} 
\usepackage[T1]{fontenc}    
\usepackage{hyperref}       
\usepackage{url}            
\usepackage{booktabs}       
\usepackage{amsfonts}       
\usepackage{nicefrac}       
\usepackage{microtype}      
\usepackage{lipsum}
\usepackage{fancyhdr}       
\usepackage{graphicx}       
\graphicspath{{media/}}     
\usepackage{amsmath}
\usepackage{longtable}
\usepackage{multirow}
\usepackage{float}
\usepackage{array, booktabs}
\usepackage{geometry}
\usepackage{graphicx}
\usepackage{algorithm2e}
\usepackage{makecell}

\title{A Machine Learning Framework for Breast Cancer Treatment Classification Using a Novel Dataset
}

\author{
  Md Nahid Hasan \\
  Department of Mathematics \\
  East Texas A\&M University \\
  Commerce, TX, USA\\
  \texttt{Nahid.Hasan@etamu.edu} \\
   \And
  Md Monzur Murshed \\
  Department of Mathematics and Statistics\\ 
  Minnesota State University, Mankato\\
  Mankato, MN USA \\
  \texttt{email@email} \\
  \AND
Md Mahadi Hasan\\
  Department of Mathematics and Statistics\\ 
  Murray State University\\
Murray, KT, USA\\ 
  \And
Faysal Ahmed Chowdhury\\
Department of Mathematics\\
Florida Gulf Coast University\\
Fort Myers, FL, USA\\
 }

\begin{document}
\maketitle

\begin{abstract}
Breast cancer (BC) remains a significant global health challenge, with personalized treatment selection complicated by the disease’s molecular and clinical heterogeneity. BC treatment decisions rely on various patient-specific clinical factors, and machine learning (ML) offers a powerful approach to predicting treatment outcomes. This study utilizes The Cancer Genome Atlas (TCGA) breast cancer clinical dataset to develop ML models for predicting the likelihood of undergoing chemotherapy or hormonal therapy. The models are trained using five-fold cross-validation and evaluated through performance metrics, including accuracy, precision, recall, specificity, sensitivity, F1-score, and area under the receiver operating characteristic curve (AUROC). Model uncertainty is assessed using bootstrap techniques, while SHAP values enhance interpretability by identifying key predictors. Among the tested models, the Gradient Boosting Machine (GBM) achieves the highest stable performance (accuracy = 0.7718, AUROC = 0.8252), followed by Extreme Gradient Boosting (XGBoost) (accuracy = 0.7557, AUROC = 0.8044) and Adaptive Boosting (AdaBoost) (accuracy = 0.7552, AUROC = 0.8016). These findings underscore the potential of ML in supporting personalized breast cancer treatment decisions through data-driven insights.
\end{abstract}

\keywords{Breast Cancer \and Machine Learning \and SHAP \and Random Forest \and AUROC \and TCGA}

\section{Introduction}
Breast cancer remains one of the most prevalent cancers worldwide, accounting for approximately 25\% of all cancer cases among women and 15\% of cancer-related deaths \cite{bray2018global}. With over 2.3 million new cases diagnosed annually, breast cancer is a major global health concern \cite{bray2020global}. According to the American Cancer Society (ACS) 2024-2025\cite{ACS2024}, breast cancer is the most common type of cancer among women in the United States, which accounts for about 30\% (or 1 in 3) of all new female cancers each year. Overall, on average, 1 out of 8 (roughly 13\%) women in the United States are at risk of developing breast cancer sometime in their life. Also, according to the ACS 2024-2025 report, breast cancer is the 2nd most leading cause of death from cancer among women. With advancements in medical science, early diagnosis and treatment options have considerably improved, yet personalized treatment selection remains a formidable challenge due to the heterogeneity of breast cancer at both the molecular and clinical levels. Properly classifying patients into appropriate treatment groups based on their clinical and pathologic features is required for improving therapeutic end results while minimizing  effects.

There are two prominent treatment options for breast cancer patients, chemotherapy and hormone therapy, each targeting distinct biological mechanisms. However, identifying the optimal treatment strategy for individual patients is a complex decision-making process influenced by patient age, tumor characteristics, hormonal receptor status, histology type, menopause status, and other clinical evaluations. While chemotherapy targets rapidly dividing cancer cells using cytotoxic agents, hormone therapy exploits the hormone receptor status, such as estrogen receptor (ER) \& progesterone receptor (PR), to suppress hormone-dependent tumor growth \cite{harbeck2019breast}. Molecular biomarkers, including ER,  PR, and human epidermal growth factor receptor-2 (HER2) status, play crucial roles in deciding whether a patient with breast cancer will respond to hormonal therapy or chemotherapy \cite{gamble2021determining}.

Among influential biomarkers for breast cancer treatment, the estrogen receptor (ER) status plays a pivotal role. About 70–80\% of breast cancers are ER-positive, meaning they contain estrogen receptors and respond favorably to hormonal treatments such as Tamoxifen. Conversely, ER-negative tumors, which lack these receptors, are typically more aggressive and require chemotherapy \cite{ACS2024}. Another significant biomarker is  HER2 status, which affects both the aggressiveness of the disease and treatment strategies. HER2-positive breast cancers, characterized by HER2 protein overexpression or gene amplification, respond well to targeted therapies such as Trastuzumab (Herceptin) and Pertuzumab. In contrast, HER2-negative cancers often require chemotherapy or alternative treatments, depending on their ER and PR status \cite{ACS2024}.

Conventional diagnostic procedures, such as mammography, ultrasound, and biopsies, often face limitations, including false positives, false negatives, and subjective interpretations \cite{jafari2018optimum}. In recent years, the emergence of machine learning (ML) as a tool for medical decision support has demonstrated great potential in improving breast cancer diagnosis, prognosis, and treatment planning. ML models are capable of processing high-dimensional data, identifying complex \& subtle patterns, and making predictions with high accuracy, hence assisting doctors in their decision-making processes. Specifically, bootstrapped-based machine learning\cite{tibshirani1993introduction} frameworks have gained traction due to their robustness in handling imbalanced datasets and reducing overfitting, which are common challenges in clinical data. Ensemble methods like bootstrap aggregating (bagging), boosting,  and random forests have shown efficacy in predicting treatment responses by leveraging diverse datasets that include patient demographics, tumor biomarkers, and clinical history. Moreover, by incorporating bootstrapping techniques, these frameworks can enhance the generalization and interpretability of predictions, which is critical in sensitive medical applications. The pathologic TNM staging system, which includes tumor size (T), lymph node involvement (N), and distant metastasis (M), remains a cornerstone in breast cancer assessment \cite{amin2017ajcc}. Emerging evidence highlights the importance of integrating additional patient- and tumor-specific features to enhance predictive accuracy. For instance, factors such as the extent of lymph node examination, tumor necrosis percentage, and anatomic subdivisions are increasingly recognized for their influence on treatment decisions and prognosis \cite{weigelt2010markers}.

Despite the advancements in ML applications for risk stratification, tumor subtype classification, and prognosis prediction, integrating clinical and pathological features to predict treatment types (chemotherapy versus hormone therapy) remains an under-explored area. Hybrid approaches combining multiple algorithms, such as Gradient Boosting Machines and Decision Trees, have shown promise in improving diagnostic outcomes \cite{zuo2023machine}. Also, feature selection methods, such as Recursive Feature Elimination (RFE), combined with Random Forests, have played a critical role in improving the interpretability of ML models \cite{gupta2018comparative}. Additionally, interpretability methods, such as SHapley Additive ExPlanations (SHAP), have proven valuable for explaining ML model predictions, fostering trust among clinicians \cite{dardouillet2022explainability}. Huang et al. \cite{huang2023increasing} used Bootstrap simulation and shapely additive explanations to evaluate variability in model metrics and covariate importance, which improves transparency and reliability in model selection.

The Cancer Genome Atlas (TCGA) \cite{weinstein2013tcga} offers a robust platform for comprehensive clinical, pathological, and molecular data on breast cancer patients. Leveraging such multi-dimensional datasets can pave the way for developing predictive frameworks to optimize therapeutic strategies \cite{Monzur2019}. Federated learning approaches, which address data privacy concerns, further enhance the applicability of ML in clinical settings \cite{sandhu2023medical}.

\begin{figure}[!ht]\label{flow}
\centering
\includegraphics[width=.8\linewidth]{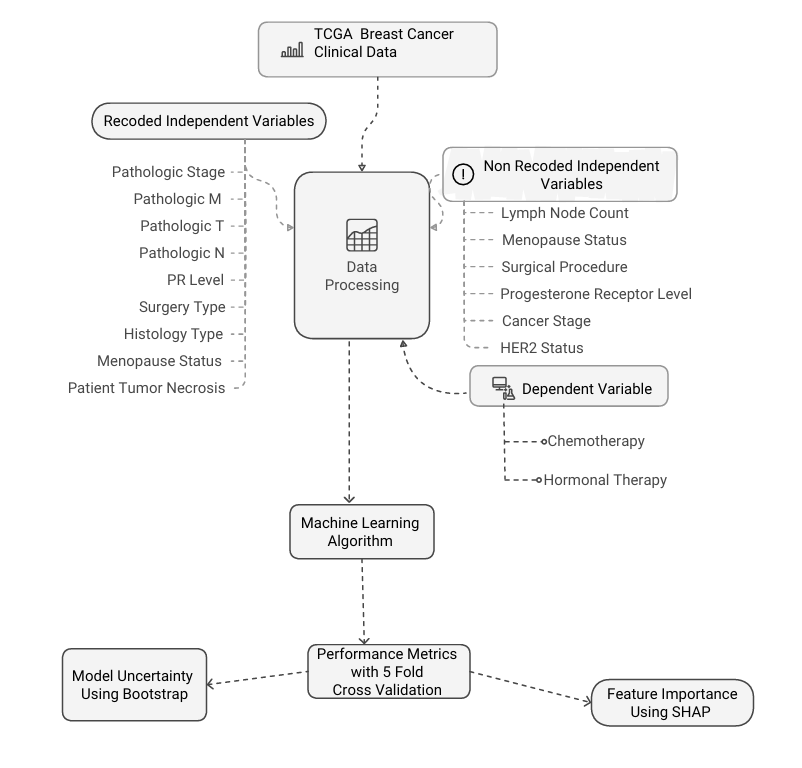}
\caption{Working Flow Chart}
\label{fig:stream}
\end{figure}

In this study, using a set of clinical and pathological features derived from the TCGA dataset, we created a novel dataset designed for breast cancer treatment classification. We trained seven well-known machine learning models on this dataset and estimated their unbiased prediction performance using five-fold cross-validation. To further assess the uncertainty of model performance, we conducted bootstrap analysis. In addition, we identified the most influential variables using SHAP analysis, enhancing model interpretability. The proposed framework aims to improve personalized treatment selection and assist clinicians in making data-driven decisions. The workflow is illustrated in Figure~\ref{fig:stream}.

In the following method section, we have described the data, independent features along with the response variable. Then we have discussed statistical analysis and model development. In the result section, we have presented bivariate analysis comprise with statistical significance of features, overall performance of ML models, feature importance using SHAP value. In the discussion section, we summarized the key findings and some limitations of this study.

\section{Methods}

\subsection{Data Description}

The dataset for this study has been obtained from The Cancer Genome Atlas (TCGA) website, TCGA is a collaboration between the National Cancer Institute (NCI) and the National Human Genome Research Institute (NHGRI). TCGA serves as a repository of molecular information from tumor and normal samples, as well as clinical details for each patient. This platform is freely accessible to researchers who can download datasets covering various types of cancers, including brain cancer and prostate cancer. The TCGA data portal contains essential information on more than 33 cancer types and is regularly updated with new datasets. In this study, we utilized the most recent data available on breast cancer.
The TCGA provides information on two types of breast cancer: ductal carcinoma and lobular carcinoma. In this research, we are considering data on both types of cancer. We have downloaded the clinical information datasets from the TCGA website ( \url{http://gdac.broadinstitute.org}).  The dataset contains information on 1,481 clinical variables collected between 2003 and 2014 and includes data from 1,198 patients.

\subsection{Dependent Variable}

TCGA data was collected over a decade, and for each patient, several visit information was recorded. Specifically, each patient underwent five follow-up visits, with detailed information recorded at each stage. In the initial phase of data processing, we focus on identifying the first treatment received by each patient, as accurately determining this can improve prognosis and help prevent cancer recurrence. After identifying the initial treatment, we remove any missing values for this variable and retain only cases where the patient received either chemotherapy or hormonal therapy—the two most common breast cancer treatments. Subsequently, after an extensive literature review, we select other relevant variables essential for breast cancer treatment classification.

\subsection{Independent Variable}

Our selection of relevant variables begins with lymph node examining count, which is assessed during the surgical procedure. A higher count indicates greater cancer spread beyond the primary tumor site. However, in our dataset, this variable contains several missing values, likely due to patients either not undergoing surgery or lacking recorded information. We interpret these missing values as cases where no lymph nodes were examined and assign them a value of zero.

Another critical predictor of treatment outcomes is menopause status. Similar to lymph node count, some values for this variable are missing. We classify these missing values as ``unknown status".
Surgical procedure type is also included in our analysis, as it plays a crucial role in treatment decisions. Missing values in this variable are assumed to indicate that no surgery was performed. Additionally, we incorporate progesterone receptor (PR) level, which provides valuable insights into tumor biology and helps determine treatment options. For this variable, missing values are replaced with ``no information”.

We also consider cancer stage, as treatment strategies vary based on disease progression. Missing values in this variable are assigned stage X to reflect the uncertainty. Lastly, HER2 immunohistochemistry receptor status is included due to its importance in guiding treatment strategies. HER2-positive cancers tend to grow more rapidly but may respond better to targeted therapies. Like the PR level, missing values for HER2 status are replaced with ``no information".

After addressing missing data, we retain 761 observations for further analysis. The next step in data processing involves recoding selected variables, as shown in Table \ref{tab:listv}.

\renewcommand{\arraystretch}{1.1} 

\small 

\begin{longtable}{p{4cm}p{10cm}} 
\caption{ List of recoded features and their categories included in the breast cancer dataset} \label{tab:listv} \\

\hline
\textbf{Variable} & \textbf{Categories} \\
\hline
\multirow{5}{4cm}{\textbf{Pathologic-stage}} 
    & Stage I: stage i, stage ia, stage ib \\
    & Stage II: stage ii, stage iia, stage iib \\
    & Stage III: stage iii, stage iiia, stage iiib, stage iiic \\
    & Stage IV: stage iv \\
    & Stage X: stage x, NA \\\hline
\multirow{3}{4cm}{\textbf{Pathologic-M}} 
    & m0: cm0 (i+), m0 \\
    & m1: m1 \\
    & m2: mx \\\hline
\multirow{4}{4cm}{\textbf{Pathologic-T}} 
    & t1: t1, t1b, t1c \\
    & t2: t2, t2a \\
    & t3: t3, t3a \\
    & t4: t4, t4b, t4d, tx \\
\hline
\multirow{4}{4cm}{\textbf{Pathologic-N}} 
    & n0: n0, n0 (i-), n0 (i+), n0 (mol+) \\
    & n1: n1, n1a, n1b, n1c, n1mi \\
    & n2: n2, n2a \\
    & n3: n3, n3a, n3b, n3c, nx \\\hline
\multirow{4}{4cm}{\textbf{PR-Level}} 
    & Low Expression: <10\% \\
    & Moderate Expression: 10-19\%, 20-29\%, 30-39\%, 40-49\% \\
    & High Expression: 50-59\%, 60-69\%, 70-79\%, 80-89\%, 90-99\% \\
    & No Information: no information \\\hline
\multirow{3}{4cm}{\textbf{Surgery-Type}} 
    & Lumpectomy: lumpectomy \\
    & Mastectomy: modified radical mastectomy, simple mastectomy \\
    & No Surgery / Other: no surgery, other \\\hline
\multirow{3}{4cm}{\textbf{Histology-Type}} 
    & Infiltrating Ductal Carcinoma: infiltrating ductal carcinoma \\
    & Infiltrating Lobular Carcinoma: infiltrating lobular carcinoma \\
    & Other Type: infiltrating carcinoma nos, medullary carcinoma,  metaplastic carcinoma, mixed histology (please specify),  mucinous carcinoma, other, specify \\\hline
\multirow{4}{4cm}{\textbf{Menopause-Status}} 
    & Pre: pre (<6 months since lmp and no prior bilateral  ovariectomy and not on estrogen replacement) \\
    & Post: post (prior bilateral ovariectomy or >12 mo since lmp with no prior hysterectomy) \\
    & Peri: peri (6-12 months since last menstrual period) \\
    & No Information/Other: status Unknown, indeterminate  (neither pre or postmenopausal) \\\hline
\multirow{4}{4cm}{\textbf{patient tumor necrosis}} 
    & No necrosis: 0 \\
    & Partial necrosis: 1-49 \\
    & Significant necrosis: 50-99 \\
    & Complete necrosis: 100 \\
\hline
\end{longtable}

We also included age, PR status, tumor nuclei percentage, and anatomic subdivision (left or right) as predictors in our study. After excluding all missing values from these variables except those in Table \ref{tab:listv} for treatment prediction, we are left with 723 observations. Of these, 467 patients received chemotherapy while the remaining patients received hormonal therapy.

\subsection{Model development}

In this study, we utilized several machine learning methods for breast cancer treatment classification based on a novel TCGA dataset. We used Adaptive boosting (Adaboost), Linear Discriment Aanalysis (LDA), Logistic Regression (LR), Random Forest (RF), Support Vetor Machines (SVM) with radial carnel, and Gradient Boosting Models (XGBoost) \cite{cox1958regression, breiman1984classification, breiman2001random, vapnik1964class, bayes1958essay, fisher1936use,  friedman2001greedy, freund1997decision}.

Initially, patients were randomly divided into five groups for cross-validation. Each group served as a test set, while the remaining four groups formed the training set. This process was repeated five times so that each group was used as a test set once. At each iteration, the training set was resampled with replacement. Models were trained on the resampled data and evaluated on the test set. This approach ensured that all observations contributed to both training and testing.

Model performance was evaluated using accuracy, area under the receiver operating characteristic curve (AUC), sensitivity, specificity, precision, and F1 score \cite{fawcett2006introduction, john2010elements}. These metrics provided a comprehensive evaluation of classification performance. To quantify prediction uncertainty, bootstrap simulation ($N = 1,000$) was performed \cite{xi2022improving}. This involved repeatedly resampling the training set, training the models, and evaluating their performance.
Finally, we computed descriptive statistics for the simulated results. These included mean, median, standard deviation, and 95\% confidence intervals, summarizing the variability in model performance. See algorithm 1.

\begin{algorithm}[h]
\caption{Machine Learning-Based Breast Cancer Treatment Classification}
\KwIn{TCGA dataset $X$}
\KwOut{Model performance metrics $V$}

Randomly split $X$ into five folds: $\{X_1, X_2, ..., X_5\}$\;
\Repeat{Bootstrap sampling reaches $N = 1000$ iterations}{
    \ForEach{fold $X_i \in \{X_1, X_2, ..., X_5\}$}{
        Set $\text{data.test} \gets X_i$, $\text{data.train} \gets X \setminus X_i$\;
        Resample $\text{data.train}$ with replacement\;
        \ForEach{model $\ell \in L = \{\text{AdaBoost, GBM, LDA, LR, RF, SVM, XGBoost}\}$}{
            Train $m$ on $\text{data.train}$\;
            Predict on $\text{data.test}$, store result $Y_{i,\ell}$\;
        }
    }
    Aggregate predictions $Y$, compute performance metrics (Accuracy, AUC, Sensitivity, Specificity, Precision, F1-score)\;
    Store results in $V$\;
}
Compute summary statistics (mean, median, standard deviation, 95\% CI for $V$)\;
Determine the best-performing model and output final results\;
\end{algorithm}

SHapley Additive exPlanations (SHAP) analysis was conducted to quantify the impact of each feature to the predictive model\cite{lundberg2017unified}. SHAP values provide a decomposition of model predictions into contributions from each feature. A summary plot was generated to show the impact of each feature on the model output. This plot provided a visual representation of each feature’s impact and variability in influencing model predictions. Mean of absolute SHAP values was computed to rank features based on their importance to the model.

All computer analyses including model training and evaluation  were performed using free statistical computing software \textbf{R} version 4.4.1 \cite{R2023}.

\section{Results}
\subsection{Bivariate analysis}
In Table \ref{tab:breast_cancer}, key features of patients undergoing chemotherapy and hormone therapy for breast cancer are compared. For this bivariate analysis, numerical variables were summarized as medians with interquartile ranges (IQR: Q1, Q3) to account for their skewed or non-normal distributions. Categorical variables were expressed as frequencies and percentages. To compare differences between chemotherapy and hormone therapy groups, chi-square tests were used for categorical variables. For continuous variables, the  Kruskal-Wallis or Mann-Whitney U test was applied to analyze skewed or non-normally distributed data. Most features in the dataset are significantly associated with the choice of treatment. However, four features—surgery type, pathologic-M, anatomic subdivision, and tumor-necrosis percent—do not show a statistically significant difference between the two groups.  Statistical significance was determined using a p-value threshold of 0.05, where p-value $<$ 0.05 indicates a significant association between a feature and treatment selection.

Table \ref{tab:breast_cancer} reveals that patients receiving hormone therapy are generally older, with a median age of 65 years, compared to 53 years for chemotherapy patients. Hormone therapy is strongly linked to positive ER (98.4\%) and PR (86.3\%) statuses, while chemotherapy is more frequently associated with negative ER (32.3\%) and PR (40.3\%) statuses. In terms of histology, Infiltrating Ductal Carcinoma is more prevalent among chemotherapy patients (74.3\%), whereas Infiltrating Lobular Carcinoma is more common in those receiving hormone therapy (27.7\%).

Chemotherapy patients tend to have more lymph nodes examined (median: 8 nodes) compared to hormone therapy patients (median: 4 nodes). Hormone therapy is predominantly used in postmenopausal patients (78.9\%), while chemotherapy is more common in premenopausal (30.2\%) and peri-menopausal (4.3\%) groups. Additionally, chemotherapy is more often associated with advanced cancer stages (Stage III: 28.1\%, Stage IV: 1.1\%), whereas hormone therapy is more frequently used for earlier stages (Stage I: 23.8\%).

Tumor size and nodal involvement also differ between the two groups. Hormone therapy patients have a higher proportion of smaller tumors (T1: 32\%), while chemotherapy patients are more likely to have larger tumors (T2 or larger: 61.5\%). Similarly, hormone therapy is more commonly prescribed to patients with no nodal involvement (N0: 57.4\%). At the same time, chemotherapy is preferred for patients with higher nodal stages. Hormone therapy is also associated with higher PR expression levels (31.6\% high expression) than chemotherapy (20.6\% high expression). Conversely, chemotherapy is more frequently used in HER2-positive cases (15.4\%) compared to hormone therapy (9\%).

In general, hormone therapy is typically administered to older, postmenopausal patients with positive ER/PR statuses, smaller tumors, and earlier stages of cancer. On the other hand, chemotherapy is more commonly associated with younger, premenopausal patients with negative ER/PR statuses, larger tumors, and more advanced cancer stages.

\setlength{\tabcolsep}{15pt}
\begin{scriptsize} 
\begin{longtable}{ccccc}
\caption{ Description of features included in the breast cancer dataset grouped by the treatment options} \label{tab:breast_cancer} \\

\toprule
\textbf{Features} & \makecell{\textbf{Chemotherapy} \\ \textbf{(N=467)}} & \makecell{\textbf{Hormone Therapy} \\ \textbf{(N=256)}} & \makecell{\textbf{Total} \\ \textbf{(N=723)}} & \textbf{p-value} \\
\midrule
\endfirsthead

\toprule
\textbf{Features} & \makecell{\textbf{Chemotherapy} \\ \textbf{(N=467)}} & \makecell{\textbf{Hormone Therapy} \\ \textbf{(N=256)}} & \makecell{\textbf{Total} \\ \textbf{(N=723)}} & \textbf{p-value} \\
\midrule
\endhead

\bottomrule
\multicolumn{5}{r}{{Continued on next page}} \\
\endfoot

\bottomrule  \multicolumn{5}{l}{*Numerical features are displayed as median ($Q1, Q3$); Categorical features are displayed as frequency (percentage)}
\endlastfoot 

\multicolumn{1}{l}{\textbf{Age}} & 53 (46, 61) & 65 (55, 74) & 57 (48, 65.5) & $<0.001$  \\ [.5em]

\multicolumn{1}{l}{\textbf{ER-Status}} & &&& $<0.001$  \\[.4em]
Negative & 151 (32.3\%) & 4 (1.6\%) & 155 (21.4\%)  & \\
Positive & 316 (67.7\%) & 252 (98.4\%) & 568 (78.6\%)  & \\[.5em]

\multicolumn{1}{l}{\textbf{PR-Status}} & &&& $<0.001$  \\[.4em]
Negative & 188 (40.3\%) & 35 (13.7\%) & 223 (30.8\%)  & \\
Positive & 279 (59.7\%) & 221 (86.3\%) & 500 (69.2\%)  & \\[.5em]

\multicolumn{1}{l}{\textbf{Surgery-Type}} & &&& 0.458  \\[.4em]
Lumpectomy & 119 (25.5\%) & 75 (29.3\%) & 194 (26.8\%)  & \\
Mastectomy & 230 (49.3\%) & 115 (44.9\%) & 345 (47.7\%)  & \\
No Surgery / Other & 118 (25.3\%) & 66 (25.8\%) & 184 (25.4\%)  & \\[.5em]

\multicolumn{1}{l}{\textbf{Histology-Type}} & &&& 0.003  \\[.4em]
Infiltrating Ductal Carcinoma & 347 (74.3\%) & 162 (63.3\%) & 509 (70.4\%)  & \\
Infiltrating Lobular Carcinoma & 80 (17.1\%) & 71 (27.7\%) & 151 (20.9\%)  & \\
Other Type & 40 (8.6\%) & 23 (9.0\%) & 63 (8.7\%)  & \\[.5em]

\multicolumn{1}{l}{\textbf{Lymph-Nodes-Examined}} & 8 (2, 17) & 4 (2, 12) & 6 (2, 15) & $<0.001$  \\[.5em]

\multicolumn{1}{l}{\textbf{Menopause-Status}} & &&& $<0.001$  \\[.4em]
No Information/Other & 35 (7.5\%) & 11 (4.3\%) & 46 (6.4\%)  & \\
Peri & 20 (4.3\%) & 6 (2.3\%) & 26 (3.6\%)  & \\
Post & 271 (58.0\%) & 202 (78.9\%) & 473 (65.4\%)  & \\
Pre & 141 (30.2\%) & 37 (14.5\%) & 178 (24.6\%)  & \\[.5em]

\multicolumn{1}{l}{\textbf{Pathologic-Stage}} & &&& $<0.001$  \\[.4em]
Stage I & 55 (11.8\%) & 61 (23.8\%) & 116 (16.0\%)  & \\
Stage II & 269 (57.6\%) & 153 (59.8\%) & 422 (58.4\%)  & \\
Stage III & 131 (28.1\%) & 35 (13.7\%) & 166 (23.0\%)  & \\
Stage IV & 5 (1.1\%) & 4 (1.6\%) & 9 (1.2\%)  & \\
Stage X & 7 (1.5\%) & 3 (1.2\%) & 10 (1.4\%)  & \\[.5em]

\multicolumn{1}{l}{\textbf{Pathologic-M}} & &&& 0.317  \\[.4em]
m0 & 393 (84.2\%) & 204 (79.7\%) & 597 (82.6\%)  & \\
m1 & 6 (1.3\%) & 4 (1.6\%) & 10 (1.4\%)  & \\
m2 & 68 (14.6\%) & 48 (18.8\%) & 116 (16.0\%)  & \\[.5em]

\multicolumn{1}{l}{\textbf{Pathologic-T}} & &&& 0.014  \\[.4em]
t1 & 105 (22.5\%) & 82 (32.0\%) & 187 (25.9\%)  & \\
t2 & 287 (61.5\%) & 139 (54.3\%) & 426 (58.9\%)  & \\
t3 & 67 (14.3\%) & 27 (10.5\%) & 94 (13.0\%)  & \\
t4 & 8 (1.7\%) & 8 (3.1\%) & 16 (2.2\%)  & \\[.5em]
                                                     
\multicolumn{1}{l}{\textbf{Pathologic-N}} & &&& $<0.001$ \\[.4em]
n0 & 186 (39.8\%) & 147 (57.4\%) & 333 (46.1\%)  & \\
n1 & 168 (36.0\%) & 79 (30.9\%) & 247 (34.2\%)  & \\
n2 & 62 (13.3\%) & 15 (5.9\%) & 77 (10.7\%)  & \\
n3 & 51 (10.9\%) & 15 (5.9\%) & 66 (9.1\%)  & \\[.5em]

\multicolumn{1}{l}{\textbf{PR-Level}} & &&& 0.002 \\[.4em]
High Expression & 96 (20.6\%) & 81 (31.6\%) & 177 (24.5\%)  & \\
Low Expression & 74 (15.8\%) & 38 (14.8\%) & 112 (15.5\%)  & \\
Moderate Expression & 37 (7.9\%) & 27 (10.5\%) & 64 (8.9\%)  & \\
No Information & 260 (55.7\%) & 110 (43.0\%) & 370 (51.2\%)  & \\[.5em]

\multicolumn{1}{l}{\textbf{Anatomic-Subdivision}} & &&& 0.589  \\[.4em]
Left & 70 (15.0\%) & 42 (16.4\%) & 112 (15.5\%)  & \\
Left Lower Inner Quadrant & 8 (1.7\%) & 6 (2.3\%) & 14 (1.9\%)  & \\
Left Lower Outer Quadrant & 21 (4.5\%) & 9 (3.5\%) & 30 (4.1\%)  & \\
Left Upper Inner Quadrant & 41 (8.8\%) & 24 (9.4\%) & 65 (9.0\%)  & \\
Left Upper Outer Quadrant & 91 (19.5\%) & 54 (21.1\%) & 145 (20.1\%)  & \\[.4em]
Right & 84 (18.0\%) & 32 (12.5\%) & 116 (16.0\%)  & \\[.4em]
Right Lower Inner Quadrant & 9 (1.9\%) & 7 (2.7\%) & 16 (2.2\%)  & \\
Right Lower Outer Quadrant & 18 (3.9\%) & 16 (6.2\%) & 34 (4.7\%)  & \\
Right Upper Inner Quadrant & 35 (7.5\%) & 22 (8.6\%) & 57 (7.9\%)  & \\
Right Upper Outer Quadrant & 90 (19.3\%) & 44 (17.2\%) & 134 (18.5\%)  & \\[.5em]

\multicolumn{1}{l}{\textbf{Tumor-Necrosis}} & &&& 0.624  \\[.4em]
No Necrosis & 263 (56.3\%) & 149 (58.2\%) & 412 (57.0\%)  & \\
Partial Necrosis & 204 (43.7\%) & 107 (41.8\%) & 311 (43.0\%)  & \\[.5em]

\multicolumn{1}{l}{\textbf{Tumor-Nuclei-Percent}} & 80 (70, 85) & 80 (70, 90) & 80 (70, 90) & 0.098  \\[.5em]

\multicolumn{1}{l}{\textbf{HER2-Status}} & &&& 0.015  \\[.4em]
Equivocal & 84 (18.0\%) & 68 (26.6\%) & 152 (21.0\%)  & \\
Indeterminate & 7 (1.5\%) & 2 (0.8\%) & 9 (1.2\%)  & \\
Negative & 267 (57.2\%) & 147 (57.4\%) & 414 (57.3\%)  & \\
No Information & 37 (7.9\%) & 16 (6.2\%) & 53 (7.3\%)  & \\
Positive & 72 (15.4\%) & 23 (9.0\%) & 95 (13.1\%)  & \\

\end{longtable}
\end{scriptsize}

\subsection{Overall performances of ML models}

\begin{figure}[ht]
\centering
\includegraphics[width=.8\linewidth]{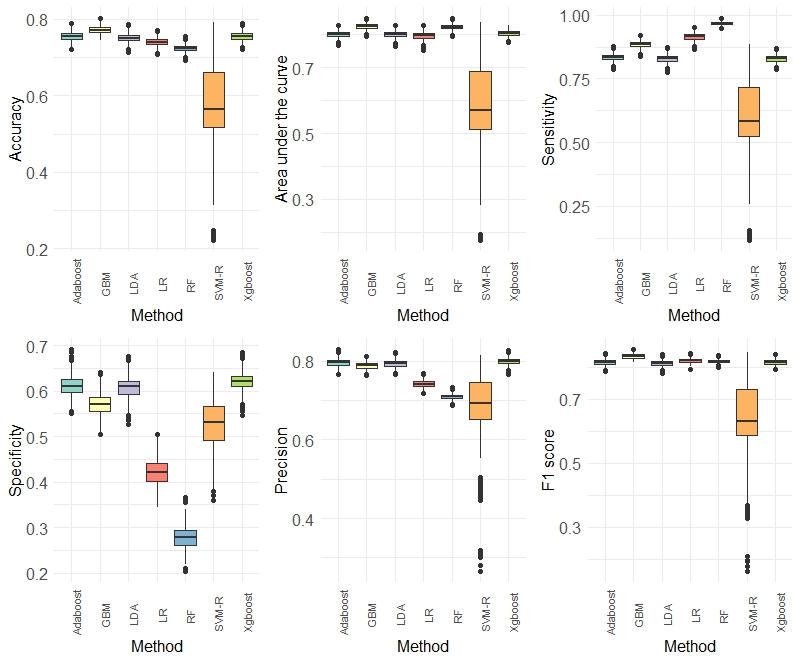}
\caption{A comparison of different machine learning methods based on various performance metrics}
\label{fig:box}
\end{figure}

We evaluated our ML models by generating 1,000 bootstrapping samples for each performance metrics. From these samples, we computed summary statistics for each performance metric. These statistics provided insights into the stability and reliability of model predictions. To further analyze the variability in performance, we constructed box plots and density plots. Box plots helped in detecting outliers, while density plots revealed the overall distribution shape of each performance metric.

The boxplot shown in Figure~\ref{fig:box} compares the performance of different machine learning methods across various performance metrics such as accuracy, AUC, sensitivity, specificity, precision, and F1 score. XGBoost, GBM, RF, and AdaBoost are the top performing models, showing a high median in most metrics and a small IQR. This is indicative of both reliability and robustness. These methods demonstrate substantial and stable performances from accuracy to AUC. In contrast, the poor performance of the SVM is reflected in its significantly lower median values and high variability, indicating low and unstable classification performance.   

XGBoost, RF, and GBM also demonstrate strong performance. Higher sensitivity and precision values confirm the ability for these three approaches to identify true positives and, hence, accurate positive predictions.
The specificity is very high for RF and XGBoost, which have been excellent in identifying true negatives, while LR and SVM-R have a poor specificity score. The F1 score confirms the balance of precision and sensitivity for XGBoost, GBM, and RF. For SVM-R, it dramatically decreases, further pointing out its poorer performance. 

Density plots show that all methods, except for SVM-R, display symmetric distributions for the performance metrics (see Supplement 1). This indicates that the model's predictions are reliable. Such alignment with normality boosts evaluation accuracy and enhances statistical inference when assessing model performance.

Table~\ref{tab:metrics} shows that GBM achieved the highest accuracy (mean = 0.7718, sd = 0.0094) and F1 score (mean = 0.8333, sd = 0.0072), indicating excellent overall balanced performance. The narrow confidence intervals (0.7712, 0.7724) suggest that this algorithm is both consistent and robust in classifying BC treatment options. AdaBoost and XGBoost closely follow GBM in terms of accuracy, sensitivity, and specificity, making them strong contenders for balanced predictions.

Among all models, RF demonstrated the best sensitivity (mean = 0.9668), meaning it excels at detecting positive cases (chemotheraphy). However, it has a very low specificity (mean = 0.2785, sd = 0.0239), resulting in a high rate of false positives. Similarly, LR exhibits high sensitivity (mean = 0.9145, sd = 0.0120) but suffers from low specificity (mean = 0.4219, sd = 0.0272), indicating poor identification of true negatives.

In contrast, SVM performed the worst overall, with the lowest accuracy (mean = 0.5714, sd = 0.1141) and AUROC (mean = 0.5842, sd = 0.1349). Additionally, its wide confidence intervals (0.5643, 0.5784) indicate high variability and instability, further confirming its weak classification ability. Even though LDA does not outperform AdaBoost and XGBoost, it delivers consistent and moderately strong results across all metrics. Its performance remains stable across different evaluation criteria. This stability makes LDA a dependable choice when prioritizing consistency over maximizing individual metrics

\begin{table}[!ht]
\centering
\caption{Performance Metrics for Various Methods}
\begin{tabular}{ccccccc}
\toprule
\textbf{Method} & \textbf{Metric} & \textbf{Mean} & \textbf{Median} & \textbf{SD} & \textbf{Lower Bound} & \textbf{Upper Bound} \\ \midrule
\textbf{Adaboost} & Accuracy & 0.7552 & 0.7552 & 0.0111 & 0.7545 & 0.7559 \\ 
 & Auroc & 0.8016 & 0.8021 & 0.0094 & 0.8011 & 0.8022 \\ 
 & Precision & 0.7967 & 0.7963 & 0.0093 & 0.7961 & 0.7973 \\ 
 & Sensitivity & 0.8339 & 0.8330 & 0.0130 & 0.8331 & 0.8347 \\ 
 & Specificity & 0.6117 & 0.6094 & 0.0218 & 0.6103 & 0.6130 \\ 
 & F1 Score & 0.8148 & 0.8148 & 0.0087 & 0.8143 & 0.8154 \\ [.5em]
\textbf{GBM} & Accuracy & 0.7718 & 0.7718 & 0.0094 & 0.7712 & 0.7724 \\ 
 & Auroc & 0.8252 & 0.8254 & 0.0081 & 0.8247 & 0.8257 \\ 
 & Precision & 0.7889 & 0.7890 & 0.0082 & 0.7884 & 0.7894 \\ 
 & Sensitivity & 0.8831 & 0.8844 & 0.0132 & 0.8823 & 0.8839 \\ 
 & Specificity & 0.5687 & 0.5703 & 0.0228 & 0.5672 & 0.5701 \\ 
 & F1 Score & 0.8333 & 0.8333 & 0.0072 & 0.8328 & 0.8337 \\ [.5em]
  
\textbf{LDA} & Accuracy & 0.7506 & 0.7510 & 0.0106 & 0.7499 & 0.7512 \\ 
 & Auroc & 0.8015 & 0.8015 & 0.0097 & 0.8009 & 0.8021 \\ 
 & Precision & 0.7940 & 0.7939 & 0.0093 & 0.7935 & 0.7946 \\ 
 & Sensitivity & 0.8290 & 0.8287 & 0.0147 & 0.8281 & 0.8299 \\ 
 & Specificity & 0.6075 & 0.6094 & 0.0233 & 0.6060 & 0.6089 \\ 
 & F1 Score & 0.8111 & 0.8114 & 0.0086 & 0.8105 & 0.8116 \\ [.5em]
\textbf{LR} & Accuracy & 0.7401 & 0.7400 & 0.0104 & 0.7394 & 0.7407 \\ 
 & Auroc & 0.7971 & 0.7977 & 0.0114 & 0.7964 & 0.7978 \\ 
 & Precision & 0.7428 & 0.7424 & 0.0086 & 0.7422 & 0.7433 \\ 
 & Sensitivity & 0.9145 & 0.9143 & 0.0120 & 0.9137 & 0.9152 \\ 
 & Specificity & 0.4219 & 0.4219 & 0.0272 & 0.4202 & 0.4236 \\ 
 & F1 Score & 0.8197 & 0.8196 & 0.0070 & 0.8192 & 0.8201 \\ [.5em]
  
\textbf{RF} & Accuracy & 0.7231 & 0.7234 & 0.0083 & 0.7226 & 0.7236 \\ 
 & Auroc & 0.8239 & 0.8242 & 0.0080 & 0.8234 & 0.8244 \\ 
 & Precision & 0.7097 & 0.7096 & 0.0066 & 0.7093 & 0.7101 \\ 
 & Sensitivity & 0.9668 & 0.9679 & 0.0064 & 0.9664 & 0.9672 \\ 
 & Specificity & 0.2785 & 0.2773 & 0.0239 & 0.2770 & 0.2800 \\ 
 & F1 Score & 0.8185 & 0.8184 & 0.0048 & 0.8182 & 0.8188 \\ [.5em]
  
\textbf{SVM-R} & Accuracy & 0.5714 & 0.5629 & 0.1141 & 0.5643 & 0.5784 \\
 & Auroc & 0.5842 & 0.5706 & 0.1349 & 0.5759 & 0.5926 \\ 
 & Precision & 0.6847 & 0.6934 & 0.0853 & 0.6794 & 0.6900 \\ 
 & Sensitivity & 0.5954 & 0.5824 & 0.1532 & 0.5859 & 0.6049 \\ 
 & Specificity & 0.5275 & 0.5313 & 0.0525 & 0.5242 & 0.5307 \\ 
 & F1 Score & 0.6335 & 0.6311 & 0.1247 & 0.6257 & 0.6412 \\ [.5em]
  
\textbf{Xgboost} & Accuracy & 0.7557 & 0.7552 & 0.0104 & 0.7551 & 0.7564 \\ 
 & Auroc & 0.8044 & 0.8046 & 0.0088 & 0.8039 & 0.8049 \\ 
 & Precision & 0.7997 & 0.7994 & 0.0088 & 0.7991 & 0.8002 \\ 
 & Sensitivity & 0.8298 & 0.8308 & 0.0126 & 0.8290 & 0.8305 \\ 
 & Specificity & 0.6207 & 0.6211 & 0.0204 & 0.6194 & 0.6220 \\ 
 & F1 Score & 0.8144 & 0.8143 & 0.0082 & 0.8139 & 0.8149 \\ \toprule
\end{tabular}
\label{tab:metrics}
\end{table}

\subsection{Feature importance using Shapley value}

\begin{figure}
    \centering
    \includegraphics[width=.45\linewidth]{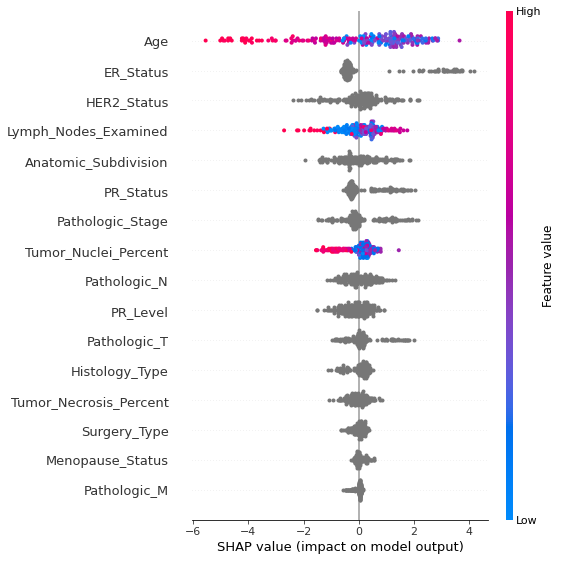}
\quad 
    \includegraphics[width=0.5\linewidth]{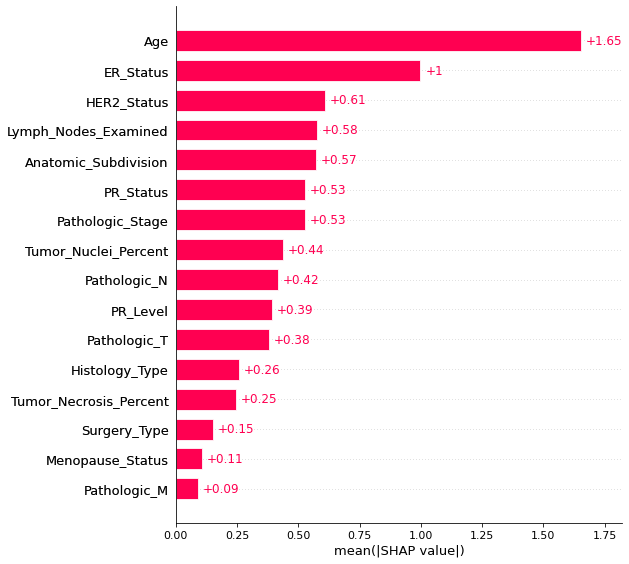}
    \caption{SHAP summary plot (left) and mean absolute SHAP value for each feature (right). $X$-axis (left) represents the change in log-odds for treatment prediction.}
    \label{fig:shap}
\end{figure}

Figure~\ref{fig:shap} presents the SHAP summary plot that provide a visual representation of the contribution of each feature to the model. $Age$ is the most influential feature with high variability, followed by $Er\_Status$ and $HER2\_Status$. The mean absolute SHAP value (in Figure~\ref{fig:shap}) ranked each feature based on their impact on the model. Higher SHAP value indicates higher contribution in predicting the model output. Higher SHAP value of age (mean = +1.65) indicates that $age$ contributed the most in predicting the treatment, followed by $Er\_Status$ (mean = +1) and $HER2\_Status$ (mean = +0.61). $Surgery\_Type$ (mean = +0.15), $menopause\_Status$ (mean = +0.11), and $Pathologis\_M$ (mean = +0.09) have very less contribution to the model output compared to $age$.

\section{Discussion}

Breast cancer is a highly treatable disease, and timely intervention significantly reduces the risk of recurrence. Chemotherapy and hormonal therapy are two primary treatment options, each with distinct mechanisms and side effects. Chemotherapy generally has a broader range of side effects, while hormonal therapy is often better tolerated but requires patients to meet specific hormonal receptor criteria. Personalized treatment selection is critical to maximizing efficacy and minimizing adverse effects.

In this study, we used machine learning methods to improve breast cancer treatment classification based on a novel dataset collected from the TCGA website. The dataset contains 723 observations of patient's demographic and clinical information. Based on our dataset, we developed seven machine-learning models to predict an appropriate treatment option (chemotherapy or hormonal therapy) for breast cancer patients. We evaluated model performance using six well-known measurements such as accuracy, AUROC, precision, sensitivity, specificity, and F1 Score.
The algorithm was designed to enhance prediction accuracy and interpretability by employing bootstrap simulation and cross-validation. This iterative approach allowed models to generalize better, reducing the risk of overfitting while estimating prediction uncertainty.

GBM produced the best results in terms of high Accuracy (0.7718) and F1 Score (0.8333) with a narrow confidence interval, which indicates this algorithm is reliable and sturdy. AdaBoost and XGBoost are the second-best models; these two models are very close to GBM in competitive metrics of Accuracy, Sensitivity, and Specificity. The RF model has high sensitivity and very low specificity. LR showed the same results as that of a RF. LDA showed comparatively fair results in all performance metrics, just slightly behind the AdaBoost and XGBoost performance. SVM showed the poorest model performance with a wide confidence interval.

The significant differences in model performance highlight the importance of selecting algorithms suited to structured clinical datasets. Ensemble-based methods, particularly GBM, XGBoost, and AdaBoost, outperform traditional classifiers in handling structured medical data due to their ability to capture nonlinear patterns and interactions among predictors.

To interpret model predictions, SHAP analysis (Figure~\ref{fig:shap}) was employed to identify influential clinical factors affecting treatment classification. The most significant predictors were Age, ER Status, and HER2 Status. These findings are consistent with previous studies that highlight the critical role of ER and HER2 status in determining treatment options \cite{carleton2022personalising}. The American Cancer Society (\url{https://www.cancer.org/cancer/types/breast-cancer/understanding-a-breast-cancer-diagnosis.html}) also highlights ER and HER2 status as key determinants in treatment selection, further validating our findings. Other factors such as Surgery Type, Menopause Status, and Pathologic M stage had comparatively lower influence on treatment classification. Understanding which features strongly influence model predictions is crucial for clinical adoption, as interpretable AI solutions are more likely to gain trust among oncologists and medical practitioners. This aligns with the broader objective of explainable AI (XAI) in healthcare, which seeks to bridge the gap between model performance and clinical utility.

Machine learning models offer evidence-based decision support by identifying complex relationships within high-dimensional datasets that may not be easily recognizable by human experts. These models can assist oncologists in making personalized treatment recommendations by integrating tumor characteristics, genetic markers, and prior treatment responses. However, several challenges remain before widespread clinical implementation, including data standardization, model interpretability, and ethical considerations regarding patient privacy.

\section{Conclusion}

This study utilizes The Cancer Genome Atlas (TCGA) breast cancer clinical dataset to develop ML models for predicting the likelihood of undergoing chemotherapy or hormonal therapy. ML models show potential in supporting personalized breast cancer treatment decisions through data-driven insights.

Despite promising results, this study has several limitations. First, only clinical variables were considered, while genetic factors play a crucial role in treatment decisions. Future models should incorporate genomic and molecular profiling data to enhance predictive accuracy. Second, the study does not account for oncologists’ expert recommendations, which are essential in real-world treatment selection. A hybrid approach incorporating both machine learning and expert-driven decision-making may yield more practical insights. Lastly, the feature set was limited, and additional factors such as treatment history, comorbidities, and lifestyle variables should be explored in future research.

In future, we aim to collaborate with healthcare practitioners to integrate genetic data and additional clinical features into our models. We also plan to conduct a systematic review to compare our findings with recent advancements in breast cancer treatment selection, ensuring that machine learning-based recommendations align with evolving medical practices.

\section*{Author contributions statement}

All author contributed equally without any conflict. 

\section*{Disclosure statement}

The authors declare no conflicts of interest.

\section*{Data availability statement}

Data is provided within the supplementary information files

\section*{Ethics Statement}

This study utilizes publicly available data from The Cancer Genome Atlas (TCGA), a project that has established comprehensive policies to address ethical, legal, and social considerations in cancer genomics research. This study aligns with TCGA's established policies and the broader ethical guidelines governing genomic research. By utilizing TCGA data, we adhere to the principles set forth to maximize public benefit while safeguarding the rights and privacy of research participants. For detailed information on TCGA's ethics and policies, please refer to the National Cancer Institute's official documentation, avaialble at \url{https://www.cancer.gov/ccg/research/genome-sequencing/tcga/history/ethics-policies}.

\bibliographystyle{unsrt}  

\begin{thebibliography}{99}

\bibitem{bray2018global} Bray, F. et al. Global cancer statistics 2018: GLOBOCAN estimates of incidence and mortality worldwide for 36 cancers in 185 countries. \emph{CA: A Cancer Journal for Clinicians}, \textbf{68}, 394--424 (2018).

\bibitem{bray2020global} Bray, F., Laversanne, M., Weiderpass, E. \& Soerjomataram, I. Global cancer statistics 2020: GLOBOCAN estimates of incidence and mortality worldwide for 36 cancers in 185 countries. \emph{CA: A Cancer Journal for Clinicians}, \textbf{70}, 313--336 (2020).

\bibitem{ACS2024} American Cancer Society. \emph{Breast Cancer Facts \& Figures 2024--2025} (2024). Accessed: 2025-01-18.

\bibitem{harbeck2019breast} Harbeck, N. et al. Breast cancer. \emph{Nature Reviews Disease Primers}, \textbf{5}, 1--31 (2019).

\bibitem{gamble2021determining} Gamble, P. et al. Determining breast cancer biomarker status and associated morphological features using deep learning. \emph{Communications Medicine}, \textbf{1}, 14 (2021).

\bibitem{jafari2018optimum} Jafari-Marandi, R., Davarzani, S., Gharibdousti, M. S. \& Smith, B. K. An optimum ANN-based breast cancer diagnosis: Bridging gaps between ANN learning and decision-making goals. \emph{Applied Soft Computing}, \textbf{72}, 108--120 (2018).

\bibitem{tibshirani1993introduction} Tibshirani, R. J. \& Efron, B. An Introduction to the Bootstrap. \emph{Monographs on Statistics and Applied Probability}, \textbf{57}, 1--436 (1993).

\bibitem{amin2017ajcc} Amin, M. et al. \emph{AJCC Cancer Staging Manual (8th ed.)}. Springer (2017).

\bibitem{weigelt2010markers} Weigelt, B., Peterse, J. \& van’t Veer, L. Breast cancer metastasis: markers and models. \emph{Nature Reviews Cancer}, \textbf{5}, 591--602 (2010).

\bibitem{zuo2023machine} Zuo, D. et al. Machine learning-based models for the prediction of breast cancer recurrence risk. \emph{BMC Medical Informatics and Decision Making}, \textbf{23}, 276 (2023).

\bibitem{gupta2018comparative} Gupta, M. \& Gupta, B. A comparative study of breast cancer diagnosis using supervised machine learning techniques. In \emph{2018 Second International Conference on Computing Methodologies and Communication (ICCMC)}, 997--1002 (IEEE, 2018).

\bibitem{dardouillet2022explainability} Dardouillet, P. et al. Explainability of image semantic segmentation through SHAP values. In \emph{International Conference on Pattern Recognition}, 188--202 (Springer, 2022).

\bibitem{huang2023increasing} Huang, A. A. \& Huang, S. Y. Increasing transparency in machine learning through bootstrap simulation and SHAP explanations. \emph{PLoS One}, \textbf{18}, e0281922 (2023).

\bibitem{weinstein2013tcga} Weinstein, J. et al. The Cancer Genome Atlas Pan-Cancer analysis project. \emph{Nature Genetics}, \textbf{45}, 1113--1120 (2013).

\bibitem{Monzur2019} Murshed, M. M. Bayesian predictive modeling for personalized treatment selection for breast cancer patients. Master’s thesis, Ball State University (2019). Restricted access.

\bibitem{sandhu2023medical} Sandhu, S. S., Gorji, H. T., Tavakolian, P., Tavakolian, K. \& Akhbardeh, A. Medical imaging applications of federated learning. \emph{Diagnostics}, \textbf{13}, 3140 (2023).

\bibitem{cox1958regression} Cox, D. R. The regression analysis of binary sequences. \emph{Journal of the Royal Statistical Society. Series B (Methodological)}, \textbf{20}, 215--232 (1958).

\bibitem{breiman1984classification} Breiman, L., Friedman, J., Stone, C. \& Olshen, R. \emph{Classification and Regression Trees}. CRC Press (1984).

\bibitem{breiman2001random} Breiman, L. Random forests. \emph{Machine Learning}, \textbf{45}, 5--32 (2001).

\bibitem{vapnik1964class} Vapnik, V. \& Chervoneva, A. On class of perceptrons. \emph{Automation and Remote Control}, \textbf{25}, 103 (1964).

\bibitem{bayes1958essay} Bayes, T. An essay towards solving a problem in the doctrine of chances. \emph{Biometrika}, \textbf{45}, 296--315 (1958).

\bibitem{fisher1936use} Fisher, R. A. The use of multiple measurements in taxonomic problems. \emph{Annals of Eugenics}, \textbf{7}, 179--188 (1936).

\bibitem{friedman2001greedy} Friedman, J. Greedy function approximation: a gradient boosting machine. \emph{Annals of Statistics}, 1189--1232 (2001).

\bibitem{freund1997decision} Freund, Y. \& Schapire, R. A decision-theoretic generalization of on-line learning and an application to boosting. \emph{Journal of Computer and System Sciences}, \textbf{55}, 119--139 (1997).

\bibitem{fawcett2006introduction} Fawcett, T. An introduction to ROC analysis. \emph{Pattern Recognition Letters}, \textbf{27}, 861--874 (2006).

\bibitem{john2010elements} Lu, J. Z. \emph{The Elements of Statistical Learning: Data Mining, Inference, and Prediction} (2010).

\bibitem{xi2022improving} Xi, N. M., Wang, L. \& Yang, C. Improving the diagnosis of thyroid cancer by machine learning and clinical data. \emph{Scientific Reports}, \textbf{12}, 11143 (2022).

\bibitem{lundberg2017unified} Lundberg, S. A unified approach to interpreting model predictions. \emph{arXiv preprint arXiv:1705.07874} (2017).

\bibitem{R2023} R Core Team. \emph{R: A Language and Environment for Statistical Computing}. R Foundation for Statistical Computing, Vienna, Austria (2024).

\bibitem{carleton2022personalising} Carleton, N. et al. Personalising therapy for early-stage oestrogen receptor-positive breast cancer in older women. \emph{The Lancet Healthy Longevity}, \textbf{3}, e54--e66 (2022).
\end{thebibliography}

\end{document}